\documentclass[fleqn,10pt]{wlscirep}
\usepackage[utf8]{inputenc}
\usepackage[T1]{fontenc}
\usepackage[section]{placeins}
\usepackage{subcaption}
\title{Ion-beam Assisted Sputtering of Titanium Nitride Thin Films}

\author[1,2]{Timothy Draher}
\author[3]{Tomas Polakovic}
\author[4]{Juliang Li}
\author[1]{Yi Li}
\author[1]{Ulrich Welp}
\author[1]{Jidong Samuel Jiang}
\author[1]{John Pearson}
\author[3]{Whitney Armstrong}
\author[3]{Zein-Eddine Meziani}
\author[4]{Clarence Chang}
\author[1]{Wai-Kwong Kwok}
\author[1,2]{Zhili Xiao}
\author[1,*]{Valentine Novosad}

\affil[1]{Argonne National Laboratory, Materials Science Division, Lemont Illinois, 60439, USA}
\affil[2]{Northern Illinois University, Department of Physics, Dekalb Illinois, 60115, USA}
\affil[3]{Argonne National Laboratory, Physics Division, Lemont Illinois, 60439, USA}
\affil[4]{Argonne National Laboratory, High Energy Physics Division, Lemont Illinois, 60439, USA}
\affil[*]{novosad@anl.gov}

\begin{abstract}
Titanium nitride is a material of interest for many superconducting devices such as nanowire microwave resonators and photon detectors. Thus, controlling the growth of TiN thin films with desirable properties is of high importance. This work aims to explore effects in ion beam-assisted sputtering (IBAS), were an observed increase in nominal critical temperature and upper critical fields are in tandem with previous work on Niobium nitride (NbN). We grow thin films of titanium nitride by both, the conventional method of DC reactive magnetron sputtering and the IBAS method, to compare their superconducting critical temperatures $T_{c}$ as functions of thickness, sheet resistance, and nitrogen flow rate. We perform electrical and structural characterizations by electric transport and x-ray diffraction measurements. Compared to the conventional method of reactive sputtering, the IBAS technique has demonstrated a 10\% increase in nominal critical temperature without noticeable variation in the lattice structure. Additionally, we explore the behavior of superconducting $T_c$ in ultra-thin films. Trends in films grown at high nitrogen concentrations follow predictions of mean-field theory in disordered films and show suppression of superconducting $T_c$ due to geometric effects, while nitride films grown at low nitrogen concentrations strongly deviate from the theoretical models.
\end{abstract}
\begin{document}

\flushbottom
\maketitle
% * <john.hammersley@gmail.com> 2015-02-09T12:07:31.197Z:
%
%  Click the title above to edit the author information and abstract
%
\thispagestyle{empty}

\section*{Introduction}

TiN has been extensively studied for its many useful mechanical, electrical, and optical properties. When fabricated into superconducting devices such as nanowire microwave resonators and photon detectors, TiN serves as an important material for fundamental structures in quantum electrical circuits, such as resonators used to multiplex large arrays of qubits~\cite{Baselmans2007}. TiN has been shown to meet the criteria desired for quantum computations and photon detection such as low RF losses at both high and low driving powers, high kinetic inductance, and tunable $T_{c}$~\cite{Wallraff2004, Vissers2010, Gao2008,Sandberg_2012, Baselmans2007, Leduc2010, Mazin2006,}. In addition, as a superconducting nitride, TiN has a high superconducting $T_{c}$, relative to elemental Ti and Ti$_{2}$N, for highly stoichiometric phases. It is a hard, mechanically robust, and stable material~\cite{Jeyachandran2007, Diserens1998, Wu1990, Vissers2013}. The composition of deposited TiN$_{x}$ compounds can be varied by changing the flux of reactive nitrogen gas present during fabrication, where varying the nitrogen concentration not only tunes the superconducting $T_{c}$, but also alters the film's crystal structure and kinetic inductance~\cite{Vissers2013, Adjaottor1995}. 

For the lowest nitrogen concentrations, an $\alpha$-Ti phase initially forms where nitrogen is incorporated interstitially. With little increase in nitrogen, there is an atomic fraction of nitrogen that forms the Ti$_{2}$N phase which is known to suppress $T_{c}$ in Ti-N compounds \cite{Greene1995}. Next, in the higher nitrogen flow regime, TiN becomes the most predominant and stable compound \cite{Spengler1978}. A mix of the TiN (111) and TiN (002) phases can form. TiN (002) is the orientation with lower surface energy and forms more elastic grains comparatively to TiN (111), however, many deposition parameters can drive the preferred growth of either orientation such as the deposition pressure, substrate bias/temperature, ion flux, and gas 
 composition\cite{Patsalas2000, Greene1995, Banerjee2002}.  Growth of TiN can be conducted using a variety of physical vapor deposition (PVD) techniques including sputtering, evaporation, and molecular beam epitaxy (MBE).

MBE allows for highly stoichiometric and ordered growth of multi-component films like TiN at low temperatures inside an ultra-high vacuum environment~\cite{Rasic2017}, while the use of reactive sputtering or evaporation promotes a more polycrystalline and amorphous lattice structure. The latter techniques offer faster growth and higher throughput at the cost of less control over crystal structure during deposition. However, sputtering and evaporation still offers the ability to grow films of high quality with desirable characteristics by tailoring the deposition parameters~\cite{Jeyachandran2007}.

In reactive DC magnetron sputtering, the target material is connected to a high power DC source that creates a plasma out of a mixture of inert gas (usually argon) and a reactive gas (in this case nitrogen) which is then confined by magnetic fields local to the source target. The gas particles are ionized by the strong electric fields and are accelerated towards the target, which knocks loose the desired sputtering atoms that then recombine with the reactive gas to form the thin film. Ion-beam assisted sputtering (IBAS) utilizes the enhanced kinematic effects of an additional ion source to bombard the sample surface during the reactive sputtering process. This effectively anneals the film surface, and promotes better adhesion~\cite{Zhang2016, Hirsch1980}. In the case of reactive IBAS, the ion-beam source also functions as the supply of the reactive gas.

In a previous work with niobium nitride, IBAS was shown to decrease the sensitivity of nitrogen to forming ideal superconducting stoichiometric films and increase $T_{c}$ ~\cite{Polakovic2018}. In this study, we aim to compare the IBAS method with conventional reactive magnetron sputtering of TiN and explore its effects on superconducting $T_{c}$, structure, and electrical properties.

\section*{Methods}
TiN films were deposited on 2-inch high resistance $(\rho > 10$~k$\Omega$cm$)$ Si (100) wafers with a thin layer of native oxide inside a commercial ultra-high vacuum sputtering system from Angstrom Engineering~\cite{Angstrom}. Two separate growth techniques were utilized at room temperature. The first being conventional DC reactive magnetron sputtering and the second with the added bombardment of nitrogen ions from a diffusive ion-beam source, adapting the IBAS method. Before deposition, the chamber vacuum was pumped down to $5 \times 10^{-9}$~Torr and the substrate surface was etched of water or organic contamination using a low energy argon ion beam. Moreover, the substrate was continuously rotated during deposition to assure uniform film growth. Samples were not heated or annealed during deposition and the temperature did not exceed 30 \textdegree C. Sputtering rates were determined by use of x-ray reflectometry and profilometer measurements on a masked twin sample.

For both methods, the chamber pressure was held fixed at 3~mTorr with a continuous mass flow of 99.9999\% argon at 30~sccm. While the reactive ultrahigh purity (99.9997\%) nitrogen gas concentration was varied from 0 up to 10 sccm. A 99.995\% titanium target was sputtered from a 3-inch diameter magnetron sputtering gun powered by a DC power source with \; P $\approx$ 11.6~W$\cdot$ cm$^{-2}$. The substrate-to-target distance sits at 5-inches with a 33\textdegree ~angle relative to the substrate surface normal. The ion-beam source was an end-Hall ion gun with attached hollow cathode for thermionic emission of electrons to neutralize the beam plasma~\cite{Kaufman1987}. It rests at a 40\textdegree ~angle from the substrate and 20\textdegree ~azimuthal from the Ti gun. During IBAS deposition, the nitrogen flow is supplied only from the ion source rather than uniformly around the substrate during conventional sputtering. Ion energies of N$_{2}$ were kept low at 100~eV to minimize any structural damage to the films and reducing the formation of microcracks that lead to pores along the surface \cite{Marchenko2008,Hibbs1984}. While maintaining a 0.5~A ion current, this is equivalent to an ion power density of 70~mW$\cdot$ cm$^{-2}$. Table ~\ref{tab:Deposition Parameters} summarizes the general deposition parameters used.

The superconducting $T_{c}$ of the TiN films was measured via a standard four-wire probe method in an ICEoxford Dry Ice cryostat and Bluefors dilution refrigerator. In addition, x-ray diffraction (XRD) analysis was conducted on films grown from both methods to determine the phase of TiN. Sheet resistance measurements followed via a four probe on a circular sample, to correct for geometric factors~\cite{yilmaz2015geometric}.

\begin{table}[h]
    \centering
    \begin{center}
    \begin{tabular}{|c|c|}
    \hline
        Parameter & Value \\
        \hline
        Base Pressure (Torr) & $5 \times 10^{-9}$\\
        Working Pressure (mTorr) & 3 \\
        Nitrogen Flow (sccm) & 0 - 10 \\
        Argon Flow (sccm) & 30 \\
        Target/Substrate Distance (inch) & 5 \\
        Deposition Temperature (°C) & $\leq$ 30 \\
        Ion Energy (eV) & 100 \\
        Ion Current (A) & 0.5 \\
        \hline
    \end{tabular}
    \end{center}
    \caption{TiN sputtering 
chamber deposition parameters.}
    \label{tab:Deposition Parameters}
\end{table}

\section*{Results}
The advantages of the IBAS method for TiN are best demonstrated by direct comparison of superconducting critical temperatures of thin films grown by conventional reactive sputtering under identical chemical conditions as the IBAS films.

\begin{figure}[h]
    \centering
    \includegraphics[width=0.8\linewidth]{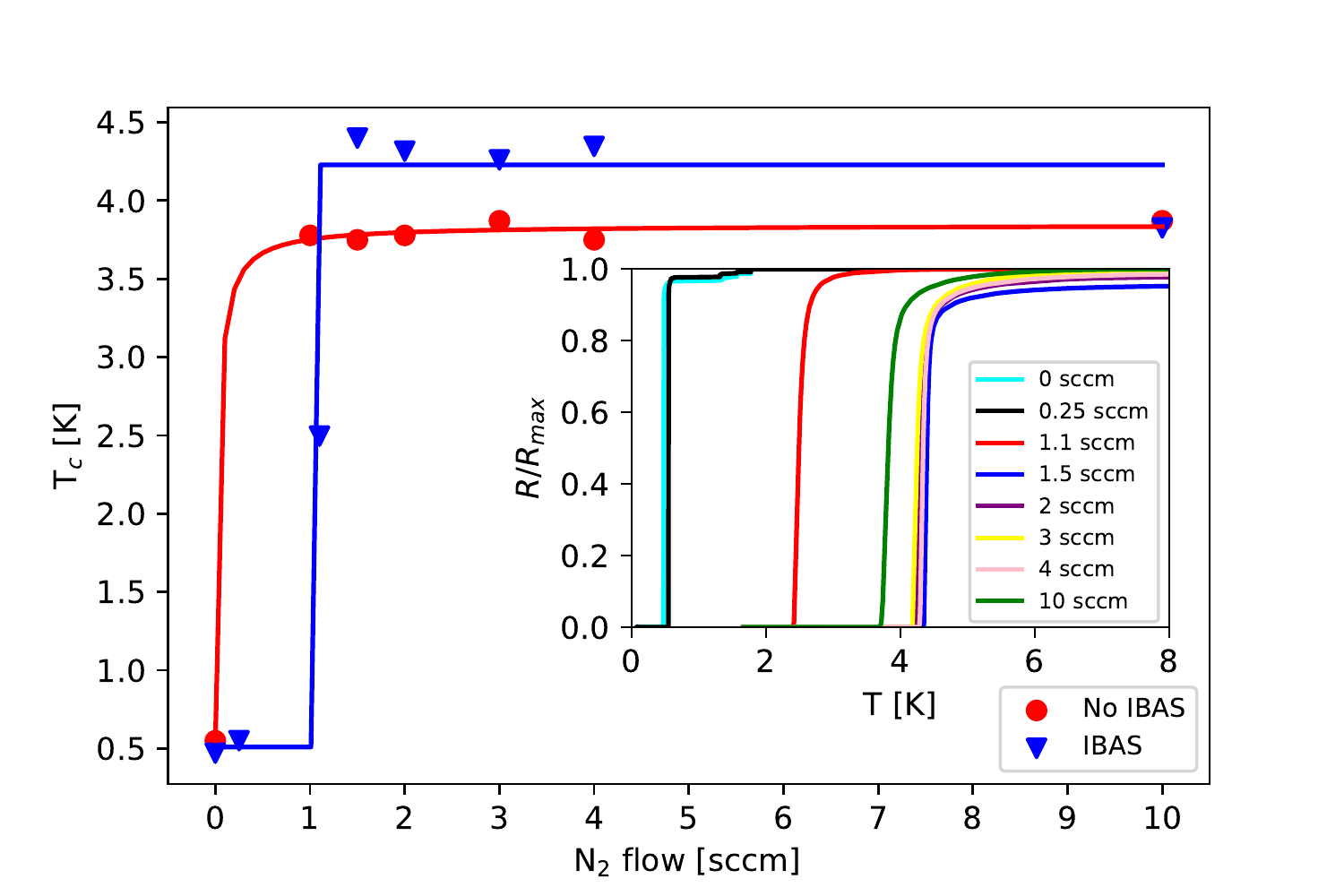}
    \caption{Comparison of bulk 300nm TiN films dependence of $T_{c}$ on nitrogen flow rates under reactive magnetron sputtering with no ion beam assist (red) and including the ion beam (blue). Trend curves are sigmoid fits meant to serve as guides to the eye. Inset: Normalized resistance transition curves for IBAS grown films at various nitrogen flow points.}
    \label{fig:Tc_comparison}
\end{figure}

The nitrogen flow dependence on superconducting $T_{c}$ for 300~nm films grown with both techniques is shown by Fig.~\ref{fig:Tc_comparison} with resistive transitions inset for IBAS grown films. While the superconducting $T_{c}$ changes little with higher nitrogen flow rates, there is large variation near the flow range of 0.5-2~sccm, where the $T_{c}$ increases sharply from 0.5 K to 4-4.5 K. The IBAS grown films show a 10\% increase in nominal superconducting $T_{c}$. The sharp increase in the superconducting $T_{c}$ is due to the formation of stoichiometric TiN as the nitrogen content increases \cite{Vissers2013}.

\begin{figure*}[h]
    \centering
    \includegraphics[width=\linewidth]{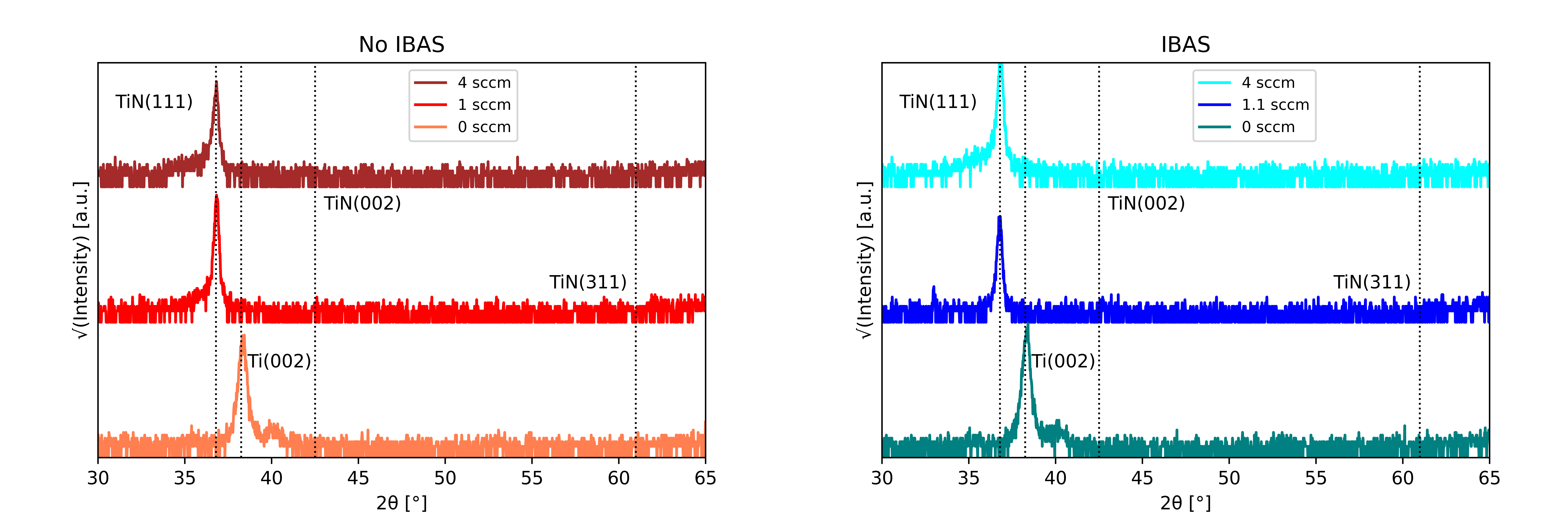}
    \caption{XRD $\omega-2\theta$ scans of 300 nm TiN films grown at room temperature via conventional reactive magnetron sputtering (left) and IBAS growth method (right) at varying nitrogen flow rates. TiN (111) is the primary orientation selected and Ti (002) is grown with no nitrogen added during deposition.}
    \label{fig:XRD}
\end{figure*}

Fig.~\ref{fig:XRD} shows comparative XRD $\omega-2\theta$ scans of films grown using both methods.     Beyond the transition where $T_{c}$ saturates in Fig.~\ref{fig:Tc_comparison}, TiN (111) is the predominant and stable phase we observe in this study \cite{Vissers2013, Jeyachandran2007}. Though it has been shown that IBAS does promote growth orientations at different ion energy and current regimes, our ion beam characteristics (100 eV ion energy and 0.5 A current) do no sufficiently promote any predominant orientation outside of TiN (111) \cite{Zhang2016, Patsalas2004, Patsalas2001}. The sputtering gas composition of argon and nitrogen used (100 - 75\% Ar and 0 - 25\% N$_{2}$) is also attributed to the preferential growth of TiN (111) \cite{Banerjee2002}. Argon's presence during sputtering promotes a  more metallic growth mode rather than nitridic. In the metallic growth mode, Ti adatoms react with nitrogen on the substrate surface \cite{Lungu1998}. However, before reacting, the Ti adatoms collect together in clusters and form low surface energy (111) planes. (111) stacks alternating layers of Ti and interstitial N resulting in rapid columnar grain growth normal to the substrate surface. The (002) orientation has the lowest surface energy and therefor the adatoms ability to diffuse outward is easier but a far slower process comparatively. Thus, the faster (111) orientation is preferred and is a factor of the limited kinetics provided by the ion source and Ar-N$_{2}$ gas composition.

The upper critical magnetic field $H_{c_{2}}$ and coherence length $\xi$ of the nominal 4 sccm IBAS film was determined by measurements of various fields perpendicular to the sample surface near $T_{c}$. From these values, the $H_{c_{2}}(T = 0)$ can be calculated by the Werthamer-Helfand-Hohenberg formula \cite{Werthamer1966}. 

\begin{equation}
    H_{c_{2}}(0) = -0.69T_{c}~\frac{dH_{c_{2}}}{dT}\bigg|_{T_{c}} 
    \label{WHH_Critical_Field}.
\end{equation}

\begin{figure*}[h]
    \centering
    \includegraphics[width=0.8\linewidth]{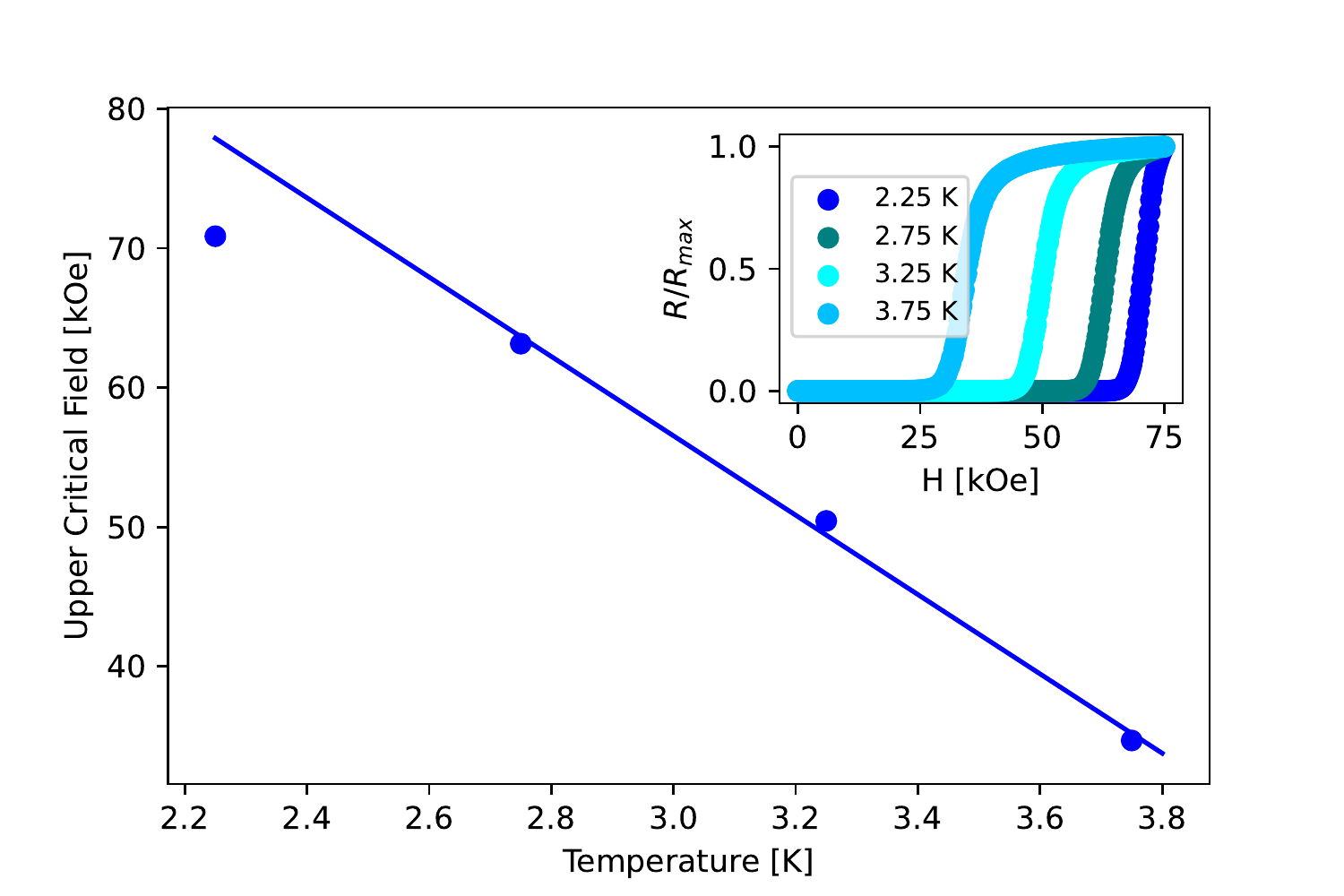}
    \caption{ Perpendicular upper critical field $H_{c_{2}}$ measured as a function of temperature for a 300 nm IBAS thin film deposited at 4 sccm. Inset: Normalized resistance measurements of the same film as a function of the applied field at temperatures close to $T_{c}$.}
    \label{fig:4sscm}
\end{figure*}

\noindent Then the in-plane Ginzburg-Landau coherence length can be calculated via,

\begin{equation}
    H_{c_{2}}(T) = \frac{\Phi_{0}}{2\pi\xi^2(T)},
    \label{Coherence_length}
\end{equation}

\noindent where $\Phi_{0}$ is the single flux quantum \cite{Tinkham1975}. Nominally this method is well founded in the region close to $T_{c}$, but in practice, it can be applied deeply into the superconducting state. Upper critical field measurement for a nominal IBAS film (300 nm at 4 sccm) can be seen in Fig.~\ref{fig:4sscm}. The calculated perpendicular critical field was found to be $H_{c_{2}}(0) = 85.4$ kOe with estimated coherence length of $\xi(0) = 1.96$ nm from fitting. The low value of $\xi(0)$ is suspected to be caused by disorder in the sputtered films, where the renormalization due to short electronic mean free path $l$ is approximately $\xi^{*} = 0.85\sqrt{\xi \times l}$~\cite{caroli1963coherence}. If we assume unperturbed $\xi = 105$ nm~\cite{faley2021titanium}, we end up with mean free path of approximately 4~\AA, which is within the reported range [3.5-7.3]~\AA ~for TiN \cite{Kardakova2015}.

Many applications of superconducting devices often require the material to be in the form of a thin film. Because of that, we also study the dependence of superconducting $T_{c}$ and its resistance on film thickness. There are many models for suppression of the superconducting state with decreasing film thickness. In this class of materials, the demise of superconducting state is often explained by either weak localization effects or electron wave leakage~\cite{Graybeal1984,Yu1976, Maekawa1982}.

\begin{figure}[h]
    \centering
    \includegraphics[width=0.8\linewidth]{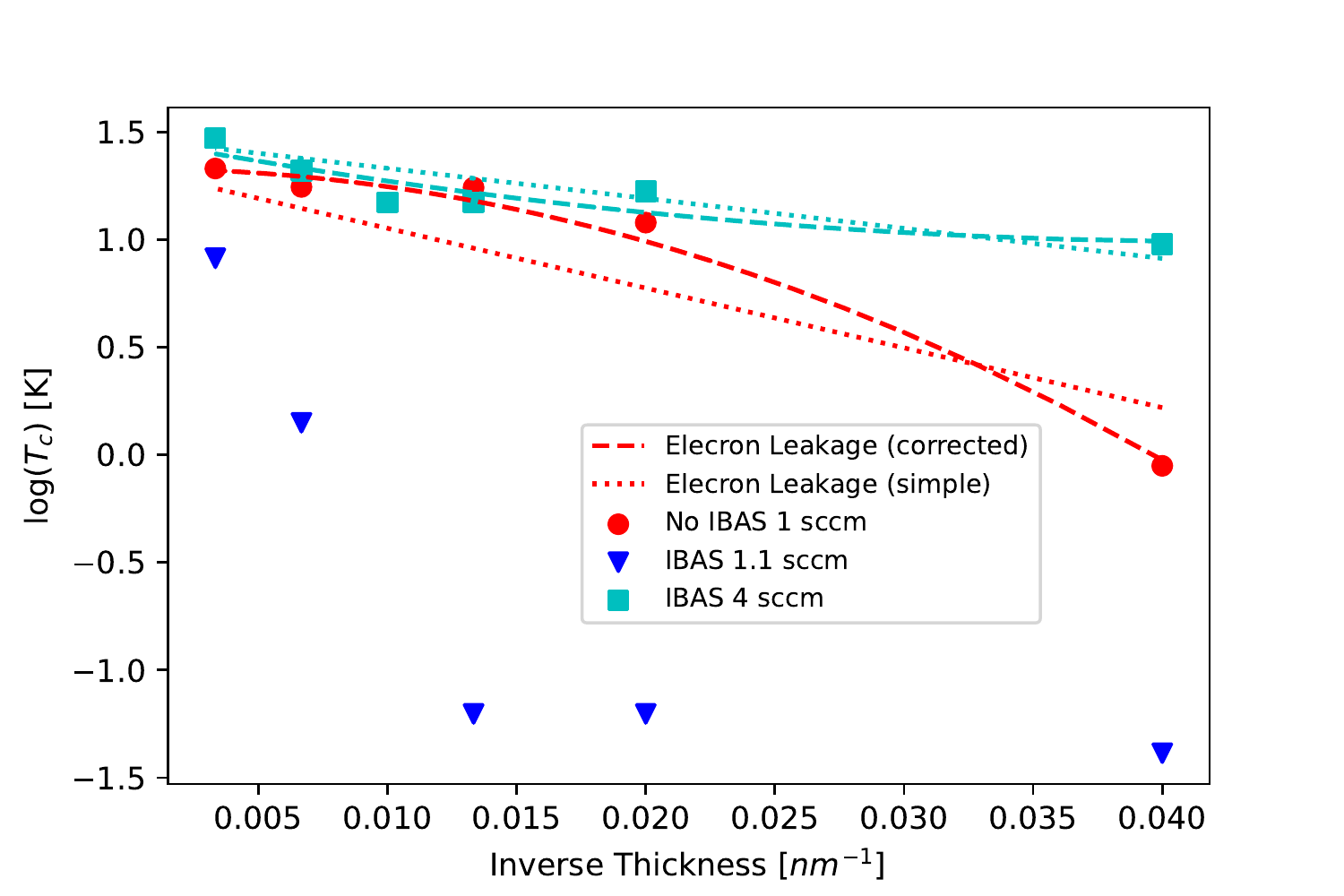}
    \caption{Dependence of superconducting $T_{c}$ on inverse film thickness consisting of different nitrogen flows. Experimental data are plotted in red circles (non-IBAS) contrasted with blue triangles and cyan squares (IBAS). Lines show best fits of different models: The dotted lines are a fit of Eq. (\ref{leakage_simple}) and the dashed lines corresponds to a fit of Eq. (\ref{leakage_corrected}). The 1.1 scc IBAS films do not follow either model.}
    \label{fig:Leakage}
\end{figure}

 The electron leakage model describes the suppression a geometric effect where the electron wave function treated as a infinite well state in the direction perpendicular to the sample surface. To preserve charge balance across the thin films, one has to nominally allow for this well to extend beyond the geometric boundary of the film, with a characteristic length as a parameter of the model.  This reduces the cooper pair density of states and suppresses $T_{c}$. This model's expression for the behavior of $T_{c}(d)$ can be shown as \cite{Yu1976}

\begin{equation}
    T_{c} = T_{c_{\infty}}\text{exp}
    \biggr[
    \frac{-b}{N(0)Vd}
    \biggr]
    \label{leakage_simple}
\end{equation}

where $T_{c_{\infty}}$ is the superconducting critical temperature of a bulk sample, $b$ is the characteristic length of electron wave leakage, and $N(0)V$ is the BCS coupling constant. By using known values for the Debye temperature $\theta_{D}$ (for TiN of 746-769~K ~\cite{Kozma2020}), and the superconducting energy gap $\Delta$ of 3~meV, we can extract the BCS coupling of N(0)V = 0.165 and use it to determine the leakage parameters in our films. Considering the disordered nature of sputtered films, one can further modify Eq. (\ref{leakage_simple}) to account for the presence of defects and film breakup with,

\begin{equation}
    T_{c} = T_{c_{\infty}}\text{exp}
    \biggr[
    \frac{-1}{N(0)V}
    \biggl(
    \frac{b}{d} + \frac{c}{d^{2}}
    \biggl)
    \biggr]
    \label{leakage_corrected}
\end{equation}

where $c$ is the parameter that accounts for film breakup and defects. Fig.~\ref{fig:Leakage} demonstrates the dependence of superconducting $T_{c}$ on film thickness $d$ for both deposition methods at nitrogen flow points of 1~sccm for the non-IBAS films, and 1.1~sccm and 4~sccm for IBAS, respectively. With application of the simpler model, we extract b = 4.58~\AA ~for the 1~sccm non-IBAS films, and b = 2.30~\AA ~for the 4~sccm IBAS films. Turning to Eq. (\ref{leakage_corrected}), a quantitatively better fit for 1 sccm non-IBAS yields values of $b = 5.97 \times 10^{-7}$~\AA ~and $c = 139.6$ \AA$^{2}$, however, these values are outside of reasonable ranges (where $b$ should be on the order of electron Fermi wavelength). For 4~sccm IBAS films, the corrected model also fails with $b = 3.72$ \AA ~and the nonphysical value of $c = -43.6$ \AA$^{2}$. Considering the almost linear trend of log($T_{c}$) for the 4~sccm films in Fig.~\ref{fig:Leakage}, it is justifiable to use the simple electron leakage theory as a valid model. The 1.1~sccm IBAS films heavily deviate from the leakage models and the suppression of superconducting $T_{c}$ has to be driven by a different mechanism.

\begin{figure}[h]
    \centering
    \includegraphics[width=0.8\linewidth]{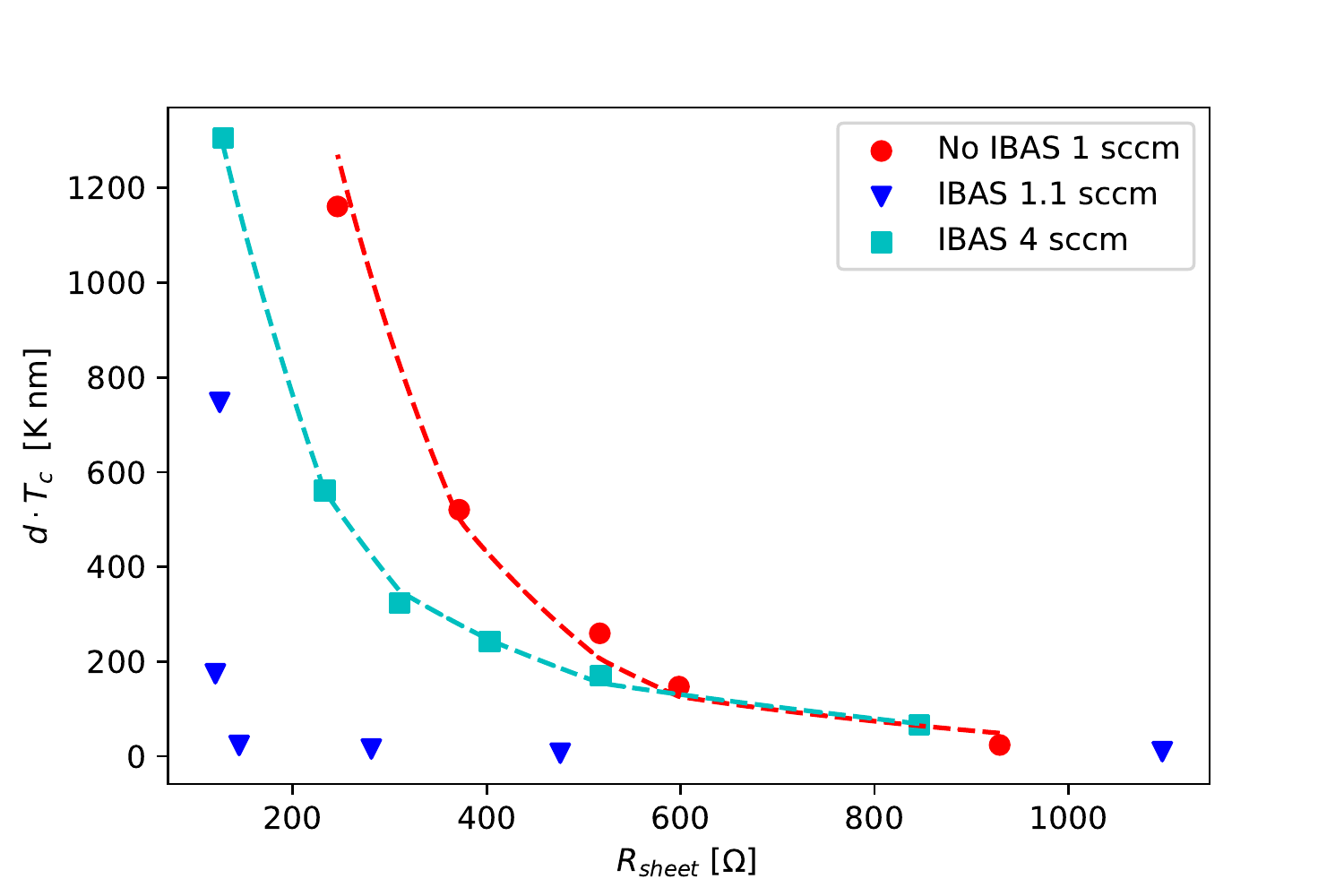}
    \caption{Dependence of superconducting $T_{c}$ scaled by thickness on sheet resistance for different nitrogen flows. Experimental data for 1 sccm non-IBAS films (red) and 1.1 (blue) and 4 sccm (cyan) IBAS films are shown respectively. The dashed lines are fits of Eq. (\ref{Ivry}) while the 1.1 sccm IBAS films heavily deviate from the model. }
    \label{fig:Ivry}
\end{figure}

Another approach to explore the suppression of superconducting $T_{c}$ is by observing the films sheet resistance on $T_{c}$. Ivry~\emph{et al.} propose an universal phenomenological power-law relationship

\begin{equation}
    d \cdot T_{c} = AR_{\text{sheet}}^{-B}
    \label{Ivry}
\end{equation}

where $R_{sheet}$ is the sheet resistance and $A$ and $B$ are fitting parameters \cite{Ivry2014}, with $B$ related to the BCS coupling $N(0)V$ in the weak coupling limit \cite{McMillan1968}. Fig. \ref{fig:Ivry} shows the scaled $T_{c}$ as a function of $R_{sheet}$. Where the 1~sccm non-IBAS films follow the Ivry fit with $B = 0.58$. The IBAS films show different behavior with decreasing nitrogen flow. The 4~sccm films still follow the power-law with $B = 0.24$ while the 1.1~sccm films again deviates largely. The $B$ values of both sets of films are lower than for films grown by atomic layer deposition gathered and reported by Ivry $B \approx [0.81-0.96]$ but within the general range of the phenomenological theory.

\begin{figure}[h]
    \centering
    \includegraphics[width=0.8\linewidth]{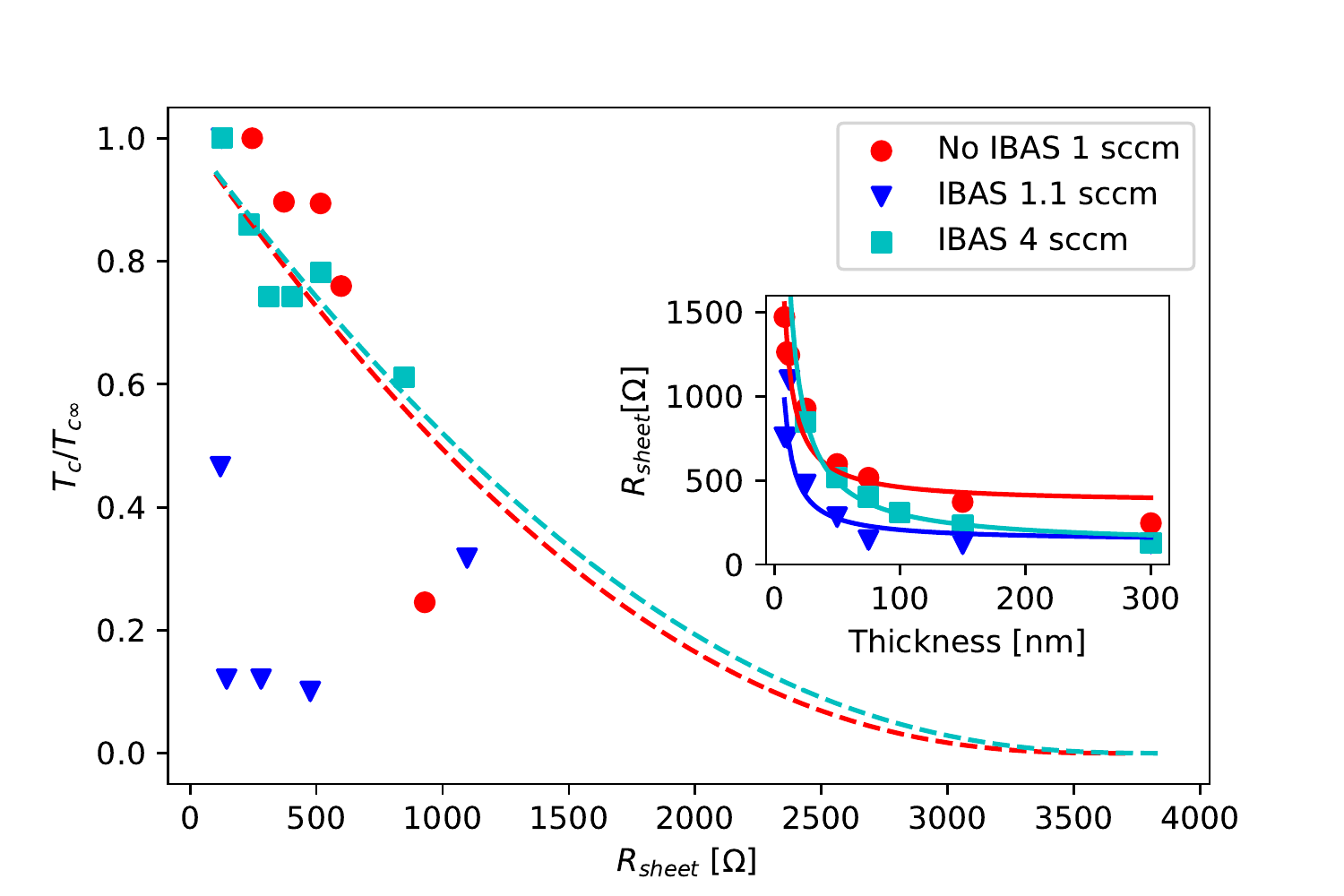}
    \caption{ Dependence of superconducting $T_{c}$ on sheet resistance. The 1 sccm non-IBAS (red), 1.1 sccm (blue) and 4 sccm (cyan) IBAS films are shown. The red and cyan curves are experimental fits of the data to Eq. (\ref{fig:Finkelstein}). The blue 1 sccm IBAS data does not follow the same model. Inset: dependence of sheet resistance on film thickness where all three curves follow a similar $\frac{1}{d}$ dependence.}
    \label{fig:Finkelstein}
\end{figure}

A more first-principles approach is to employ results of renormalization methods to further attempt to explain the suppression of $T_{c}$~\cite{Finkel'stein1994}:

\begin{equation}
    \frac{T_{c}}{T_{c_{\infty}}} = \text{exp}
    \biggr[
    \frac{-1}{\gamma}
    \biggr]
    \times 
    \Biggr[
    \frac{1+\frac{\sqrt{r/2}}{\gamma - r/4}}{1-\frac{\sqrt{r/2}}{\gamma - r/4}}
    \Biggr]^{1/\sqrt{2r}}
    \label{Finkelstein}
\end{equation}

where $\gamma = \frac{1}{log(k_{b}T_{c_{\infty}}\tau/\hbar)}$ and $r = \frac{R_{sheet}}{(2\pi^{2}\hbar/e^{2})}$, $k_{b}$ is the Boltzmann constant, e is the elementary charge, and $\tau$ is the electron elastic scattering time. Fig. \ref{fig:Finkelstein} shows the approximately linear behavior of the 1~sccm non-IBAS films with $\tau = 1.59 \times 10^{-15}$s and similarly $\tau = 1.63 \times 10^{-15}$s for the 4~sccm IBAS set. Notably, the 1.1~sccm IBAS films again stray away from the model. This further suggests that within the nitrogen flow transition region, where $T_{c}$ rises rapidly (Fig. \ref{fig:Tc_comparison}), the IBAS process has a substantial effect on the mesoscale structure of the films.

A potential mechanism driving the suppression of $T_{c}$ and increase of resistivity is due to lattice point defects such as oxygen substitutions and nitrogen vacancies, which are possible in the nitrogen-poor environment at low N$_{2}$ flows \cite{Ohya2014}. The ion beam increases adatom mobility by imparting increased momentum of incident nitrogen ions onto the substrate, which increases the probability that such defects could occur, especially around this critical nitrogen flow point, where formation of stoichiometric TiN begins to form. In addition, we conducted atomic force microscopy imaging that revealed little difference in morphologies between the 1 and 4 sccm samples for both growth methods. The only difference we could identify is that the grain sizes were larger for the 4 sccm samples, as expected\cite{Banerjee2002}.

\section*{Discussion}
In this study, we have investigated the superconducting properties of TiN thin films grown by two different methods: DC reactive sputtering and ion-beam assisted sputtering (IBAS) at room temperature. Our results have shown that the IBAS method has several benefits when compared to the DC reactive sputtering method. Specifically, we have observed a 10\% higher nominal critical temperature using the IBAS method.

The behavior of the superconducting critical temperature $T_{c}$ in films grown at high nitrogen concentrations follows the predictions of the electron leakage model and mean-field theory for disordered thin films. These models suggest that a higher nitrogen concentration promotes a more uniform film with fewer non-superconducting interfacial layers, leading to an increase in $T_{c}$. However, in the lower nitrogen flow regime, the experimental data deviates heavily from these models. We observed a non-monotonous trend in $T_{c}$ as a function of thickness and resistivity, which is an effect that requires further exploration and deserves a separate future study.

\section*{Data Availability}

The datasets generated during and/or analysed during the current study are available from the corresponding author on reasonable request.

\bibliography{sample}

\begin{thebibliography}{10}
\urlstyle{rm}
\expandafter\ifx\csname url\endcsname\relax
  \def\url#1{\texttt{#1}}\fi
\expandafter\ifx\csname urlprefix\endcsname\relax\def\urlprefix{URL }\fi
\expandafter\ifx\csname doiprefix\endcsname\relax\def\doiprefix{DOI: }\fi
\providecommand{\bibinfo}[2]{#2}
\providecommand{\eprint}[2][]{\url{#2}}

\bibitem{Baselmans2007}
\bibinfo{author}{Baselmans, J. J.~A.} \emph{et~al.}
\newblock \bibinfo{journal}{\bibinfo{title}{Development of high-q
  superconducting resonators for use as kinetic inductance detectors}}.
\newblock {\emph{\JournalTitle{Advances in Space Research}}}
  \textbf{\bibinfo{volume}{40}}, \bibinfo{pages}{708--713},
  \doiprefix\url{https://doi.org/10.1016/j.asr.2007.06.041}
  (\bibinfo{year}{2007}).

\bibitem{Wallraff2004}
\bibinfo{author}{Wallraff, A.} \emph{et~al.}
\newblock \bibinfo{journal}{\bibinfo{title}{Strong coupling of a single photon
  to a superconducting qubit using circuit quantum electrodynamics}}.
\newblock {\emph{\JournalTitle{Nature}}} \textbf{\bibinfo{volume}{431}},
  \bibinfo{pages}{162--167} (\bibinfo{year}{2004}).

\bibitem{Vissers2010}
\bibinfo{author}{Vissers, M.~R.} \emph{et~al.}
\newblock \bibinfo{journal}{\bibinfo{title}{Low loss superconducting titanium
  nitride coplanar waveguide resonators}}.
\newblock {\emph{\JournalTitle{Applied Physics Letters}}}
  \textbf{\bibinfo{volume}{97}}, \bibinfo{pages}{232509},
  \doiprefix\url{10.1063/1.3517252} (\bibinfo{year}{2010}).

\bibitem{Gao2008}
\bibinfo{author}{Gao, J.} \emph{et~al.}
\newblock \bibinfo{journal}{\bibinfo{title}{Experimental evidence for a surface
  distribution of two-level systems in superconducting lithographed microwave
  resonators}}.
\newblock {\emph{\JournalTitle{Applied Physics Letters}}}
  \textbf{\bibinfo{volume}{92}}, \bibinfo{pages}{152505},
  \doiprefix\url{10.1063/1.2906373} (\bibinfo{year}{2008}).

\bibitem{Sandberg_2012}
\bibinfo{author}{Sandberg, M.} \emph{et~al.}
\newblock \bibinfo{journal}{\bibinfo{title}{Etch induced microwave losses in
  titanium nitride superconducting resonators}}.
\newblock {\emph{\JournalTitle{Applied Physics Letters}}}
  \textbf{\bibinfo{volume}{100}}, \bibinfo{pages}{262605},
  \doiprefix\url{10.1063/1.4729623} (\bibinfo{year}{2012}).

\bibitem{Leduc2010}
\bibinfo{author}{Leduc, H.} \emph{et~al.}
\newblock \bibinfo{journal}{\bibinfo{title}{Titanium nitride films for
  ultrasensitive microresonator detectors}}.
\newblock {\emph{\JournalTitle{Applied Physics Letters}}}
  \textbf{\bibinfo{volume}{97}}, \doiprefix\url{10.1063/1.3480420}
  (\bibinfo{year}{2010}).

\bibitem{Mazin2006}
\bibinfo{author}{Mazin, B.~A.} \emph{et~al.}
\newblock \bibinfo{journal}{\bibinfo{title}{Position sensitive x-ray
  spectrophotometer using microwave kinetic inductance detectors}}.
\newblock {\emph{\JournalTitle{Applied Physics Letters}}}
  \textbf{\bibinfo{volume}{89}}, \bibinfo{pages}{222507},
  \doiprefix\url{10.1063/1.2390664} (\bibinfo{year}{2006}).

\bibitem{}
\bibinfo{author}{Bretz-Sullivan, T.~M.} \emph{et~al.}
\newblock \bibinfo{journal}{\bibinfo{title}{High kinetic inductance nbtin
  superconducting transmission line resonators in the very thin film limit}}.
\newblock {\emph{\JournalTitle{Applied Physics Letters}}}
  \textbf{\bibinfo{volume}{121}}, \bibinfo{pages}{52602},
  \doiprefix\url{10.1063/5.0100961} (\bibinfo{year}{2022}).

\bibitem{Jeyachandran2007}
\bibinfo{author}{Jeyachandran, Y.~L.}, \bibinfo{author}{Narayandass, S.},
  \bibinfo{author}{Mangalaraj, D.}, \bibinfo{author}{Areva, S.} \&
  \bibinfo{author}{Mielczarski, J.~A.}
\newblock \bibinfo{journal}{\bibinfo{title}{Properties of titanium nitride
  films prepared by direct current magnetron sputtering}}.
\newblock {\emph{\JournalTitle{Materials Science and Engineering: A}}}
  \textbf{\bibinfo{volume}{445-446}}, \bibinfo{pages}{223--236},
  \doiprefix\url{https://doi.org/10.1016/j.msea.2006.09.021}
  (\bibinfo{year}{2007}).

\bibitem{Diserens1998}
\bibinfo{author}{Diserens, M.}, \bibinfo{author}{Patscheider, J.} \&
  \bibinfo{author}{Lévy, F.}
\newblock \bibinfo{journal}{\bibinfo{title}{Improving the properties of
  titanium nitride by incorporation of silicon}}.
\newblock {\emph{\JournalTitle{Surface and Coatings Technology}}}
  \textbf{\bibinfo{volume}{108-109}}, \bibinfo{pages}{241--246},
  \doiprefix\url{https://doi.org/10.1016/S0257-8972(98)00560-X}
  (\bibinfo{year}{1998}).

\bibitem{Wu1990}
\bibinfo{author}{Wu, H.~Z.} \emph{et~al.}
\newblock \bibinfo{journal}{\bibinfo{title}{Characterization of titanium
  nitride thin films}}.
\newblock {\emph{\JournalTitle{Thin Solid Films}}}
  \textbf{\bibinfo{volume}{191}}, \bibinfo{pages}{55--67},
  \doiprefix\url{https://doi.org/10.1016/0040-6090(90)90274-H}
  (\bibinfo{year}{1990}).

\bibitem{Vissers2013}
\bibinfo{author}{Vissers, M.~R.} \emph{et~al.}
\newblock \bibinfo{journal}{\bibinfo{title}{Characterization and in-situ
  monitoring of sub-stoichiometric adjustable superconducting critical
  temperature titanium nitride growth}}.
\newblock {\emph{\JournalTitle{Thin Solid Films}}}
  \textbf{\bibinfo{volume}{548}}, \bibinfo{pages}{485--488},
  \doiprefix\url{https://doi.org/10.1016/j.tsf.2013.07.046}
  (\bibinfo{year}{2013}).

\bibitem{Adjaottor1995}
\bibinfo{author}{Adjaottor, A.}, \bibinfo{author}{Meletis, E.},
  \bibinfo{author}{Logothetidis, S.}, \bibinfo{author}{Alexandrou, I.} \&
  \bibinfo{author}{Kokkou, S.}
\newblock \bibinfo{journal}{\bibinfo{title}{Effect of substrate bias on
  sputter-deposited ticx, tiny and ticxny thin films}}.
\newblock {\emph{\JournalTitle{Surface and Coatings Technology}}}
  \textbf{\bibinfo{volume}{76-77}}, \bibinfo{pages}{142--148},
  \doiprefix\url{10.1016/0257-8972(95)02594-4} (\bibinfo{year}{1995}).

\bibitem{Greene1995}
\bibinfo{author}{Greene, J.~E.}, \bibinfo{author}{Sundgren, J.},
  \bibinfo{author}{Hultman, L.}, \bibinfo{author}{Petrov, I.} \&
  \bibinfo{author}{Bergstrom, D.~B.}
\newblock \bibinfo{journal}{\bibinfo{title}{Development of preferred
  orientation in polycrystalline tin layers grown by ultrahigh vacuum reactive
  magnetron sputtering}}.
\newblock {\emph{\JournalTitle{Applied Physics Letters}}}
  \textbf{\bibinfo{volume}{67}}, \bibinfo{pages}{2928--2930},
  \doiprefix\url{10.1063/1.114845} (\bibinfo{year}{1995}).

\bibitem{Spengler1978}
\bibinfo{author}{Spengler, W.}, \bibinfo{author}{Kaiser, R.},
  \bibinfo{author}{Christensen, A.~N.} \& \bibinfo{author}{Müller-Vogt, G.}
\newblock \bibinfo{journal}{\bibinfo{title}{Raman scattering,
  superconductivity, and phonon density of states of stoichiometric and
  nonstoichiometric tin}}.
\newblock {\emph{\JournalTitle{Physical Review B}}}
  \textbf{\bibinfo{volume}{17}}, \bibinfo{pages}{1095--1101},
  \doiprefix\url{10.1103/PhysRevB.17.1095} (\bibinfo{year}{1978}).

\bibitem{Patsalas2000}
\bibinfo{author}{Patsalas, P.}, \bibinfo{author}{Charitidis, C.} \&
  \bibinfo{author}{Logothetidis, S.}
\newblock \bibinfo{journal}{\bibinfo{title}{The effect of substrate temperature
  and biasing on the mechanical properties and structure of sputtered titanium
  nitride thin films}}.
\newblock {\emph{\JournalTitle{Surface and Coatings Technology}}}
  \textbf{\bibinfo{volume}{125}}, \bibinfo{pages}{335--340},
  \doiprefix\url{10.1016/S0257-8972(99)00606-4} (\bibinfo{year}{2000}).

\bibitem{Banerjee2002}
\bibinfo{author}{Banerjee, R.}, \bibinfo{author}{Chandra, R.} \&
  \bibinfo{author}{Ayyub, P.}
\newblock \bibinfo{journal}{\bibinfo{title}{Influence of the sputtering gas on
  the preferred orientation of nanocrystalline titanium nitride thin films}}.
\newblock {\emph{\JournalTitle{Thin Solid Films}}}
  \textbf{\bibinfo{volume}{405}}, \bibinfo{pages}{64--72},
  \doiprefix\url{10.1016/S0040-6090(01)01705-9} (\bibinfo{year}{2002}).

\bibitem{Rasic2017}
\bibinfo{author}{Rasic, D.}, \bibinfo{author}{Sachan, R.},
  \bibinfo{author}{Chisholm, M.~F.}, \bibinfo{author}{Prater, J.} \&
  \bibinfo{author}{Narayan, J.}
\newblock \bibinfo{journal}{\bibinfo{title}{Room temperature growth of
  epitaxial titanium nitride films by pulsed laser deposition}}.
\newblock {\emph{\JournalTitle{Crystal Growth \& Design}}}
  \textbf{\bibinfo{volume}{17}}, \bibinfo{pages}{6634--6640},
  \doiprefix\url{10.1021/acs.cgd.7b01278} (\bibinfo{year}{2017}).
\newblock \bibinfo{note}{Doi: 10.1021/acs.cgd.7b01278}.

\bibitem{Zhang2016}
\bibinfo{author}{Zhang, L.}, \bibinfo{author}{Tong, S.}, \bibinfo{author}{Liu,
  H.}, \bibinfo{author}{Li, Y.} \& \bibinfo{author}{Wang, Z.}
\newblock \bibinfo{journal}{\bibinfo{title}{Effects of sputtering and assisting
  ions on the orientation of titanium nitride films fabricated by ion beam
  assisted sputtering deposition from metal target}}.
\newblock {\emph{\JournalTitle{Materials Letters}}}
  \textbf{\bibinfo{volume}{171}}, \bibinfo{pages}{304--307},
  \doiprefix\url{https://doi.org/10.1016/j.matlet.2016.02.100}
  (\bibinfo{year}{2016}).

\bibitem{Hirsch1980}
\bibinfo{author}{Hirsch, E.} \& \bibinfo{author}{Varga, I.}
\newblock \bibinfo{journal}{\bibinfo{title}{Thin film annealing by ion
  bombardment}}.
\newblock {\emph{\JournalTitle{Thin Solid Films}}}
  \textbf{\bibinfo{volume}{69}}, \bibinfo{pages}{99--105},
  \doiprefix\url{10.1016/0040-6090(80)90207-2} (\bibinfo{year}{1980}).

\bibitem{Polakovic2018}
\bibinfo{author}{Polakovic, T.} \emph{et~al.}
\newblock \bibinfo{journal}{\bibinfo{title}{Room temperature deposition of
  superconducting niobium nitride films by ion beam assisted sputtering}}.
\newblock {\emph{\JournalTitle{APL Materials}}} \textbf{\bibinfo{volume}{6}},
  \bibinfo{pages}{76107}, \doiprefix\url{10.1063/1.5031904}
  (\bibinfo{year}{2018}).

\bibitem{Angstrom}
\bibinfo{howpublished}{See https://angstromengineering.com/products/evovac/ for
  details corresponding to the sputtering system.}

\bibitem{Kaufman1987}
\bibinfo{author}{Kaufman, H.~R.}, \bibinfo{author}{Robinson, R.~S.} \&
  \bibinfo{author}{Seddon, R.~I.}
\newblock \bibinfo{journal}{\bibinfo{title}{End‐hall ion source}}.
\newblock {\emph{\JournalTitle{Journal of Vacuum Science \& Technology A:
  Vacuum, Surfaces, and Films}}} \textbf{\bibinfo{volume}{5}},
  \bibinfo{pages}{2081--2084}, \doiprefix\url{10.1116/1.574924}
  (\bibinfo{year}{1987}).

\bibitem{Marchenko2008}
\bibinfo{author}{Marchenko, I.~G.} \& \bibinfo{author}{Neklyudov, I.~M.}
\newblock \bibinfo{journal}{\bibinfo{title}{Film nanostructure formation during
  low-temperature pvd deposition using partially ionized atomic fluxes}}.
\newblock {\emph{\JournalTitle{Journal of Physics: Conference Series}}}
  \textbf{\bibinfo{volume}{113}}, \bibinfo{pages}{012014},
  \doiprefix\url{10.1088/1742-6596/113/1/012014} (\bibinfo{year}{2008}).

\bibitem{Hibbs1984}
\bibinfo{author}{Hibbs, M.}, \bibinfo{author}{Johansson, B.},
  \bibinfo{author}{Sundgren, J.-E.} \& \bibinfo{author}{Helmersson, U.}
\newblock \bibinfo{journal}{\bibinfo{title}{Effects of substrate temperature
  and substrate material on the structure of reactively sputtered tin films}}.
\newblock {\emph{\JournalTitle{Thin Solid Films}}}
  \textbf{\bibinfo{volume}{122}}, \bibinfo{pages}{115--129},
  \doiprefix\url{10.1016/0040-6090(84)90003-8} (\bibinfo{year}{1984}).

\bibitem{yilmaz2015geometric}
\bibinfo{author}{Yilmaz, S.}
\newblock \bibinfo{journal}{\bibinfo{title}{The geometric resistivity
  correction factor for several geometrical samples}}.
\newblock {\emph{\JournalTitle{Journal of Semiconductors}}}
  \textbf{\bibinfo{volume}{36}}, \bibinfo{pages}{082001}
  (\bibinfo{year}{2015}).

\bibitem{Patsalas2004}
\bibinfo{author}{Patsalas, P.}, \bibinfo{author}{Gravalidis, C.} \&
  \bibinfo{author}{Logothetidis, S.}
\newblock \bibinfo{journal}{\bibinfo{title}{Surface kinetics and subplantation
  phenomena affecting the texture, morphology, stress, and growth evolution of
  titanium nitride films}}.
\newblock {\emph{\JournalTitle{Journal of Applied Physics}}}
  \textbf{\bibinfo{volume}{96}}, \bibinfo{pages}{6234--6246},
  \doiprefix\url{10.1063/1.1811389} (\bibinfo{year}{2004}).

\bibitem{Patsalas2001}
\bibinfo{author}{Patsalas, P.} \& \bibinfo{author}{Logothetidis, S.}
\newblock \bibinfo{journal}{\bibinfo{title}{Optical, electronic, and transport
  properties of nanocrystalline titanium nitride thin films}}.
\newblock {\emph{\JournalTitle{Journal of Applied Physics}}}
  \textbf{\bibinfo{volume}{90}}, \bibinfo{pages}{4725--4734},
  \doiprefix\url{10.1063/1.1403677} (\bibinfo{year}{2001}).

\bibitem{Lungu1998}
\bibinfo{author}{Lungu, C.}, \bibinfo{author}{Futsuhara, M.},
  \bibinfo{author}{Takai, O.}, \bibinfo{author}{Braic, M.} \&
  \bibinfo{author}{Musa, G.}
\newblock \bibinfo{journal}{\bibinfo{title}{Noble gas influence on reactive
  radio frequency magnetron sputter deposition of tin films}}.
\newblock {\emph{\JournalTitle{Vacuum}}} \textbf{\bibinfo{volume}{51}},
  \bibinfo{pages}{635--640}, \doiprefix\url{10.1016/S0042-207X(98)00264-4}
  (\bibinfo{year}{1998}).

\bibitem{Werthamer1966}
\bibinfo{author}{Werthamer, N.~R.}, \bibinfo{author}{Helfand, E.} \&
  \bibinfo{author}{Hohenberg, P.~C.}
\newblock \bibinfo{journal}{\bibinfo{title}{Temperature and purity dependence
  of the superconducting critical field, $\uppercase{H}_{c_{2}}$. iii. electron
  spin and spin-orbit effects}}.
\newblock {\emph{\JournalTitle{Physical Review}}}
  \textbf{\bibinfo{volume}{147}}, \bibinfo{pages}{295--302},
  \doiprefix\url{10.1103/PhysRev.147.295} (\bibinfo{year}{1966}).

\bibitem{Tinkham1975}
\bibinfo{author}{Tinkham, M.}
\newblock \emph{\bibinfo{title}{Introduction to Superconductivity / Michael
  Tinkham}} (\bibinfo{publisher}{McGraw-Hill}, \bibinfo{year}{1975}).

\bibitem{caroli1963coherence}
\bibinfo{author}{Caroli, C.}, \bibinfo{author}{De~Gennes, P.} \&
  \bibinfo{author}{Matricon, J.}
\newblock \bibinfo{journal}{\bibinfo{title}{Coherence length and penetration
  depth of dirty superconductors}}.
\newblock {\emph{\JournalTitle{Physik der kondensierten Materie}}}
  \textbf{\bibinfo{volume}{1}}, \bibinfo{pages}{176--190}
  (\bibinfo{year}{1963}).

\bibitem{faley2021titanium}
\bibinfo{author}{Faley, M.~I.}, \bibinfo{author}{Liu, Y.} \&
  \bibinfo{author}{Dunin-Borkowski, R.~E.}
\newblock \bibinfo{journal}{\bibinfo{title}{Titanium nitride as a new
  prospective material for nanosquids and superconducting nanobridge
  electronics}}.
\newblock {\emph{\JournalTitle{Nanomaterials}}} \textbf{\bibinfo{volume}{11}},
  \bibinfo{pages}{466} (\bibinfo{year}{2021}).

\bibitem{Kardakova2015}
\bibinfo{author}{Kardakova, A.~I.} \emph{et~al.}
\newblock \bibinfo{journal}{\bibinfo{title}{Electron–phonon energy relaxation
  time in thin strongly disordered titanium nitride films}}.
\newblock {\emph{\JournalTitle{IEEE Transactions on Applied
  Superconductivity}}} \textbf{\bibinfo{volume}{25}}, \bibinfo{pages}{1--4},
  \doiprefix\url{10.1109/TASC.2014.2364516} (\bibinfo{year}{2015}).

\bibitem{Graybeal1984}
\bibinfo{author}{Graybeal, J.~M.} \& \bibinfo{author}{Beasley, M.~R.}
\newblock \bibinfo{journal}{\bibinfo{title}{Localization and interaction
  effects in ultrathin amorphous superconducting films}}.
\newblock {\emph{\JournalTitle{Physical Review B}}}
  \textbf{\bibinfo{volume}{29}}, \bibinfo{pages}{4167--4169},
  \doiprefix\url{10.1103/PhysRevB.29.4167} (\bibinfo{year}{1984}).

\bibitem{Yu1976}
\bibinfo{author}{Yu, M.}, \bibinfo{author}{Strongin, M.} \&
  \bibinfo{author}{Paskin, A.}
\newblock \bibinfo{journal}{\bibinfo{title}{Consistent calculation of boundary
  effects in thin superconducting films}}.
\newblock {\emph{\JournalTitle{Physical Review B}}}
  \textbf{\bibinfo{volume}{14}}, \bibinfo{pages}{996--1001},
  \doiprefix\url{10.1103/PhysRevB.14.996} (\bibinfo{year}{1976}).

\bibitem{Maekawa1982}
\bibinfo{author}{Maekawa, S.} \& \bibinfo{author}{Fukuyama, H.}
\newblock \bibinfo{journal}{\bibinfo{title}{Localization effects in
  two-dimensional superconductors}}.
\newblock {\emph{\JournalTitle{Journal of the Physical Society of Japan}}}
  \textbf{\bibinfo{volume}{51}}, \bibinfo{pages}{1380--1385},
  \doiprefix\url{10.1143/JPSJ.51.1380} (\bibinfo{year}{1982}).

\bibitem{Kozma2020}
\bibinfo{author}{Kozma, A.}
\newblock \bibinfo{journal}{\bibinfo{title}{Thermodynamic, thermal and elastic
  properties of titanium nitride tin: Comparison of various data and
  determination of the most reliable values}}.
\newblock {\emph{\JournalTitle{Technology transfer: fundamental principles and
  innovative technical solutions}}} \textbf{\bibinfo{volume}{4}},
  \bibinfo{pages}{14--17}, \doiprefix\url{10.21303/2585-6847.2020.001475}
  (\bibinfo{year}{2020}).

\bibitem{Ivry2014}
\bibinfo{author}{Ivry, Y.} \emph{et~al.}
\newblock \bibinfo{journal}{\bibinfo{title}{Universal scaling of the critical
  temperature for thin films near the superconducting-to-insulating
  transition}}.
\newblock {\emph{\JournalTitle{Physical Review B}}}
  \textbf{\bibinfo{volume}{90}}, \bibinfo{pages}{214515},
  \doiprefix\url{10.1103/PhysRevB.90.214515} (\bibinfo{year}{2014}).

\bibitem{McMillan1968}
\bibinfo{author}{McMillan, W.~L.}
\newblock \bibinfo{journal}{\bibinfo{title}{Transition temperature of
  strong-coupled superconductors}}.
\newblock {\emph{\JournalTitle{Physical Review}}}
  \textbf{\bibinfo{volume}{167}}, \bibinfo{pages}{331--344},
  \doiprefix\url{10.1103/PhysRev.167.331} (\bibinfo{year}{1968}).

\bibitem{Finkel'stein1994}
\bibinfo{author}{Finkel'stein, A.~M.}
\newblock \bibinfo{journal}{\bibinfo{title}{Suppression of superconductivity in
  homogeneously disordered systems}}.
\newblock {\emph{\JournalTitle{Physica B: Condensed Matter}}}
  \textbf{\bibinfo{volume}{197}}, \bibinfo{pages}{636--648},
  \doiprefix\url{https://doi.org/10.1016/0921-4526(94)90267-4}
  (\bibinfo{year}{1994}).

\bibitem{Ohya2014}
\bibinfo{author}{Ohya, S.} \emph{et~al.}
\newblock \bibinfo{journal}{\bibinfo{title}{Room temperature deposition of
  sputtered tin films for superconducting coplanar waveguide resonators}}.
\newblock {\emph{\JournalTitle{Superconductor Science and Technology}}}
  \textbf{\bibinfo{volume}{27}}, \bibinfo{pages}{015009},
  \doiprefix\url{10.1088/0953-2048/27/1/015009} (\bibinfo{year}{2014}).

\end{thebibliography}

\section*{Acknowledgements}

This work was supported by the U. S. Department of Energy (DOE), Office of Science, Offices of Nuclear Physics, Basic Energy Sciences, Materials Sciences and Engineering Division under Contract \#~DE-AC02-06CH11357.

\section*{Author contributions statement}

Thin film growth and transport measurements conducted by T.D. Dilution refrigerator transport measurements conducted by J.L. Manuscript and analysis of results due to T.D. and T.P. All authors reviewed the manuscript.

\section*{Competing Interests}

The authors declare no competing interests.

\section*{Additional information}

\textbf{Correspondence} and requests for materials should be addressed to V.N.

\end{document}